\documentstyle[preprint,aps]{revtex}

\tightenlines

\begin{document}

\title{Zero temperature black holes in semiclassical gravity}

\author{Paul R. Anderson\cite{PRA}}

\address{Department of Physics, Wake Forest University,
Winston-Salem, NC 27109, USA}

\author{William A.\ Hiscock\cite{BH}}

\address{Department of Physics, Montana State University, Bozeman,
MT 59717, USA}

\author{Brett E.\ Taylor\cite{BTA}}

\address{Department of Chemistry and Physics,
Radford University, Radford, VA 24142, USA} 

\maketitle

\begin{abstract}
The semiclassical Einstein equations are solved to first order in
$\epsilon = \hbar/M^2$ for the case of an extreme or nearly extreme Reissner-Nordstr\"{o}m
black hole perturbed by the vacuum stress-energy of quantized
free fields.  It is shown that, for realistic fields of spin $0$, $1/2$, or $1$,
any zero temperature black hole solution to the equations must have
an event horizon at $r_h < |Q|$, with $Q$ the charge of the black hole.
It is further shown that no black hole solutions with $r_h < |Q|$ can
be obtained by solving the semiclassical Einstein equations perturbatively.
\end{abstract}

Static spherically symmetric zero temperature black holes have
proven to be very interesting and important at the classical,
semiclassical, and quantum levels. Classically the only static
spherically symmetric black hole solution to Einstein's equations
with zero surface gravity (and hence zero temperature) is the
extreme Reissner-Nordstr\"{o}m (ERN) black hole, which possesses
a charge equal in magnitude to its mass. At the quantum level,
the statistical mechanical entropy of zero temperature (extreme)
black holes has been calculated in string
theory\cite{strominger:1996} and shown to be identical to the
usual Bekenstein-Hawking formula for the thermodynamic entropy.
The usual semiclassical temperature and entropy calculations for
ERN black holes have all been made in the test field
approximation where the effects of quantized fields on the
spacetime geometry are not considered. However, it is well known
that quantum effects alter the spacetime geometry near the event
horizon of a black hole.  In particular they can change its
surface gravity and hence its
temperature\cite{bardeen:1981,York,THA}.

In a previous paper\cite{paper1} we examined the semiclassical
backreaction due to the vacuum stress-energy of massless and
massive free quantized fields with spin $0$, $1/2$, and $1$ on a
static Reissner-Nordstr\"{o}m (RN) black hole.
We  calculated the first-order (in $\hbar$) corrections to
the RN background and found, in all cases examined, that the surface gravity was
nonzero because the energy density of the quantized field on the horizon is
negative, $\langle T_t^t \rangle > 0$.\cite{MTW}
This led us to conclude that macroscopic zero temperature
black holes might not exist in semiclassical gravity.
After this work was published a paper appeared by Lowe\cite{lowe}
which finds a zero temperature solution to the linearized semiclassical
backreaction equations.

In this paper we investigate the perturbation series more generally than
we did previously\cite{paper1}.  We find that for nearly extreme black holes,
the surface gravity (and
temperature) of the black hole decreases as the radius of the event horizon
$r_h$ decreases.  Thus these solutions form a sequence and the sequence
appears to terminate 
at the zero temperature solution found by Lowe\cite{lowe}. In our previous
paper the lowest temperature solution examined had 
$r_h = |Q|$ and $ \kappa = 4 \pi |Q| \langle T_t^t \rangle > 0$;  
Lowe's proposed solution has
$r_h = |Q|(1-4 \pi Q^2 \langle T_t^t \rangle) < |Q|$ and $\kappa = 0$.
We discuss the validity of the solutions in this sequence and show that 
it is not possible to use perturbation
theory to obtain those with $r_h < |Q|$, including the one proposed by Lowe.
However it is possible that they
nevertheless approximate exact solutions to the full nonlinear semiclassical
equations.  We also discuss a potential problem that may occur for all
zero temperature black hole solutions to the semiclassical backreaction
equations.  

The general static spherically symmetric metric can be written in
the form\cite{MTW}:
\begin{equation}
        ds^2 = - f(r)dt^2 + h(r) dr^2 + r^2 d\Omega^2 \; ,
\label{gen_fh}
\end{equation}
where $d\Omega^2$ is the metric of the two-sphere. The metric can
describe a black hole with an event horizon at $r = r_h$ if
$f(r_h) = 0$.  To avoid having a scalar curvature singularity at
the event horizon it is necessary that $h^{-1}(r_h) = 0$ as well. 
The surface gravity of such a black hole is
\begin{equation}
        \kappa = \left(\frac{1}{2}\right)\frac{f'}{\sqrt{f h}}
                \Biggm| _{r = r_h} \; ,
\label{kappa}
\end{equation}
where the prime represents a derivative with respect to $r$ and
the expression is evaluated at the horizon radius, $r_h$. The
temperature is then $T = \kappa/(2 \pi)$.

Since we wish to perturb the spacetime with the vacuum energy of
quantized fields, we begin by considering the general
Reissner-Nordstr\"{o}m metric as the ``bare'' state.  For the RN
metric,
\begin{equation}
    f(r) = h^{-1}(r) = 1 -  \frac{2M}{r} + \frac{Q^2}{r^2} \;,
\label{rn_metric}
\end{equation}
where $Q$ is the electric charge and $M$ is the mass of the black
hole. The outer event horizon is located at
\begin{equation}
        r_{+} = M + \sqrt{M^2-Q^2} \; .
\label{rplus}
\end{equation}
For the ERN black hole $\vert Q \vert = M$.

In semiclassical gravity, the geometry is treated classically
while the matter fields are quantized.  In examining the
semiclassical perturbations of the RN metric caused by the vacuum
energy of quantized fields, we continue to treat the background
electromagnetic field as a classical field. The right hand side
of the semiclassical Einstein equations will then contain both
classical and quantum stress-energy contributions,
\begin{equation}
         G_{\mu\nu} = 8 \pi \left[\left(T_{\mu\nu}\right)^C
         +  \langle  T_{\mu\nu} \rangle  \right] \; .
\label{QEFE2}
\end{equation}
We consider the situation where the black hole is in thermal
equilibrium (whether at zero or nonzero temperature) with the
quantized field; the perturbed geometry then continues to be
static and spherically symmetric.  To first order in $\epsilon =
\hbar/M^2$ the general form of the perturbed RN metric may be
written as:
\begin{equation}
        ds^2 = -[1 + 2\epsilon \rho(r) ]
                \left(1 - \frac{2m(r)}{r} + \frac{Q^2}{r^2}\right) dt^2
                + \left(1 - \frac{2m(r)}{r} +
                \frac{Q^2}{r^2}\right)^{-1} dr^2 + r^2 d\Omega ^2 \; ,
\label{pertrn}
\end{equation}
The function $m(r)$ contains both the classical mass and a
first-order quantum perturbation,
\begin{equation}
 m(r) = M[1 + \epsilon \mu (r)] \; .
\label{mofr}
\end{equation}
The metric perturbation functions, $\rho (r)$ and $\mu (r)$, are
determined by solving the semiclassical Einstein equations
expanded to first order in $\epsilon$,
\begin{eqnarray}
        \frac{d \mu}{dr} &=& - \frac{4\pi r^2}{M \epsilon}
                \langle T^t\!_t \rangle  \; ,
\label{mu_eqn} \\
        \frac{d \rho}{d r} &=& \frac{4 \pi r}{\epsilon}
                \left(1 - \frac{2M}{r} + \frac{Q^2}{r^2} \right)^{-1}
                \left[\langle T^r\!_r \rangle - \langle T^t\!_t
                \rangle \right]
                \; .
\label{rho_eqn}
\end{eqnarray}

Assuming the perturbation expansion remains valid, the functions
$\mu(r)$ and $\rho(r)$, obtained by integrating Eqs.\ (\ref
{mu_eqn}-\ref{rho_eqn}), will contain constants of integration.
It is convenient to define them as the values of the metric
perturbations on the unperturbed horizon at $r_+$, so that $\mu(r_+) = C_1$
and $\rho(r_+) = C_2$.  Since we are working in perturbation theory,
the values of these quantities on
the actual horizon are, to leading order, also $C_1$ and $C_2$ respectively.
Then to first order in $\epsilon$ the value
of $m(r)$ at the horizon is 
$m(r_h) = M(1+\epsilon C_1)$.  It is clear that
$C_1$ represents a finite renormalization of the mass $M$ of the
black hole. As in previous work\cite{York,THA}, we hereafter
denote the renormalized perturbed mass at the horizon,
 $m(r_h) = M(1+\epsilon C_1)$, by $M_R$.

The location of the
horizon is given by the vanishing of $f(r_h)$ and $h^{-1}(r_h)$.
This condition is equivalent to
\begin{equation}
r_h^2 - 2 M_R r_h + Q^2 = 0 \;.
\label{eq:horizon}
\end{equation}
Thus for a given value of $r_h$
\begin{equation}
M_R = \frac{1}{2} \left(r_h + \frac{Q^2}{r_h} \right) \;.
\label{eq:mhorizon}
\end{equation}
There exists a sequence of black hole solutions which have
$M_R = |Q| + O(\epsilon^2)$.  For solutions in this sequence
$r_h = |Q|(1 + \epsilon\, b)$ with $b$ a constant of order unity
or less.   

The surface gravity for these solutions to first order in $\epsilon$ is
\begin{equation}
\kappa =  \frac{\epsilon}{|Q|} \; [b - |Q| \mu'(|Q|)] \ .
\label{eq:kappa1}
\end{equation}
Thus as the value of $r_h$ decreases for solutions in this sequence
the surface gravity does also.  The sequence terminates with a zero
temperature black hole at
\begin{eqnarray}
b &=& |Q| \, \epsilon \mu'(r_+) \;. 
\end{eqnarray}
It is the zero temperature solution found by Lowe\cite{lowe}. 
 From Eq.\ (\ref{mu_eqn}) one sees that $\mu'(|Q|)$ is proportional to
the energy density of the quantized fields evaluated at $r=|Q|$ in the background
spacetime.  Since the energy density of these fields in an ERN background spacetime
 is negative at $r=|Q|$ in
all physically realistic cases for free fields of spin 0, 1/2, and 1\cite{paper1}, 
it is clear that if a zero temperature black hole 
solution is to exist then $b < 0$.
In our previous paper\cite{paper1} we considered solutions with $b \ge 0$.

There is a problem for the solutions with $b < 0$, corresponding to $r_h < |Q|$,
 including the zero temperature solution found by
Lowe. In a perturbation expansion the stress-energy tensor of the
quantized fields is computed in the background spacetime, and is only known for
values $r \ge r_{+}$, where $r_{+}$ is the radius of the
event horizon in the background (unperturbed) spacetime.  But the event horizons of
these proposed perturbative solutions lie inside the event horizons of the unperturbed
RN or ERN black holes, $ r_h < |Q| \leq r_{+}$. In particular for the ERN black hole
there is absolutely no justification to extrapolating the stress-energy values
inside $r_{+}$ because the event horizon is also a Cauchy horizon. Thus
perturbation theory cannot be applied in these cases.

It is still conceivable that these solutions do represent a valid set of
approximate solutions to the full nonlinear semiclassical equations. To establish this,
however, it would be necessary to compute $\langle T_{\mu}^{\nu} \rangle$ in the
semiclassical spacetime with $r_h < |Q|$ and check the self-consistency of the solution.

There is another potential problem that could occur for any proposed semiclassical
solution with $\kappa = T = 0$. It has been shown by Trivedi\cite{trivedi}
that in two spacetime dimensions the stress-energy tensor for
the quantized scalar field diverges on the event horizons
of most zero-temperature black holes including the (2-d) ERN black hole.
In four dimensions two of us\cite{AHL} have shown numerically that no such divergence
exists for massless quantized scalar fields in the ERN background.
However, it is quite possible that such a divergence could occur on the event
horizon of the perturbed zero temperature black hole.  

\section*{Acknowledgments}
P.\ R.\ A.\ would like to thank G.\ Cook, W.\ Kerr and O.\ Zaslavskii for helpful conversations.
This work was supported in part by National Science Foundation
Grant No. PHY-9734834 at Montana State University and Nos.
PHY-9800971 and PHY-0070981 at Wake Forest University.


\begin{references}

\bibitem[*]{PRA}
electronic mail address: anderson@wfu.edu

\bibitem[\dagger]{BH}
electronic mail address:  hiscock@physics.montana.edu

\bibitem[\ddagger]{BTA}
electronic mail address:  brett@peloton.runet.edu


\bibitem{strominger:1996} A.\ Strominger and C.\ Vafa, Phys. Lett. {\bf
B379}, 99 (1996).

\bibitem{bardeen:1981} J.\ M.\ Bardeen, Phys.\ Rev.\ Lett.\ {\bf 46},
382 (1981).

\bibitem{York} J.\ W.\ York,  Jr., Phys.\ Rev.\ D {\bf 31}, 775 (1985);
         D.\ Hochberg, T.\ W.\ Kephart, and J.\ W.\ York, Jr.,
        Phys.\ Rev.\ D {\bf 48}, 479 (1993);
         P.\ R.\ Anderson, W.\ A.\ Hiscock, J.\ Whitesell, and
        J.\ W.\ York Jr., Phys.\ Rev.\ D {\bf 50}, 6427 (1994).


\bibitem{THA} B.\ E.\ Taylor, W.\ A.\ Hiscock, and P.\ R.\ Anderson,
        Phys.\ Rev.\ D.\ {\bf 61}, 84021 (2000).

\bibitem{paper1} P.\ R.\ Anderson, W.\ A.\ Hiscock, and B.\ E.\ Taylor,
        Phys.\ Rev.\ Lett.\ {\bf 85}, 2438 (2000).

\bibitem{MTW}Throughout we use units such that $\hbar = c = G = k_B =
1$.  Our
        conventions are those of C.\ W.\ Misner, K.\ S.\ Thorne, and J.\
A.\ Wheeler,
        {\it Gravitation } (Freeman, San Francisco, 1973).


\bibitem{lowe} D.\ A.\ Lowe, Preprint, gr-qc/0011053.


\bibitem{AHL} P.\ R.\ Anderson, W.\ A.\ Hiscock, and D.\ J.\ Loranz,
        Phys.\ Rev.\ Lett.\ {\bf 74}, 4365 (1995).

\bibitem{trivedi} S.\ P.\ Trivedi, Phys.\ Rev.\ D{\bf 47}, 4233 (1993).

\end{references}
\end{document}